\documentclass[fleqn,usenatbib]{mnras}

\usepackage[T1]{fontenc}
\usepackage{ae,aecompl}

\usepackage{epsfig,graphicx}	
\usepackage{amsmath}	
\usepackage{amssymb}	
\usepackage{natbib}
\usepackage{comment}
\usepackage{dcolumn}
\usepackage{txfonts}
\usepackage{times}
\usepackage{bm}

\title[Disc-Jet Coupling in EXO 1745-248]{Disc-Jet Coupling in the Terzan 5 Neutron Star X-ray Binary EXO 1745$-$248}

\author[A.J. Tetarenko et al.]{A.J. Tetarenko,$^{1}$\thanks{E-mail: tetarenk@ualberta.ca}
A. Bahramian,$^{1}$
G.R. Sivakoff,$^{1}$
E. Tremou,$^{2}$
M. Linares,$^{3,4,5}$
V. Tudor,$^{6}$
\newauthor
J.C.A Miller-Jones,$^{6}$
C.O. Heinke,$^{1}$
L. Chomiuk,$^{2}$
J. Strader,$^{2}$
D. Altamirano,$^{7}$
N. Degenaar,$^{8,9}$
\newauthor
T. Maccarone,$^{10}$
A. Patruno,$^{11,12}$
A. Sanna,$^{13}$
and R. Wijnands$^{9}$
\\
$^{1}$Department of Physics, University of Alberta, CCIS 4-181, Edmonton, AB T6G 2E1, Canada\\
$^{2}$Department of Physics and Astronomy, Michigan State University, East Lansing, MI 48824, USA\\
$^{3}$Instituto de Astrof\'{i}sica de Canarias, c/ V\'{i}a L\'{a}ctea s/n, E-38205 La Laguna, Tenerife, Spain\\
$^{4}$Departamento de Astrof\'{i}sica, Universidad de la Laguna, La Laguna, E-38205, S/C de Tenerife, Spain\\
$^{5}$Institutt for fysikk, NTNU, Trondheim, Norway\\
$^{6}$International Centre for Radio Astronomy Research- Curtin University, GPO Box U1987, Perth, WA 6845, Australia\\
$^{7}$Physics and Astronomy, University of Southampton, Southampton, Hampshire SO17 1BJ, UK\\
$^{8}$Institute of Astronomy, University of Cambridge, Madingley Road, Cambridge CB3 OHA, UK\\
$^{9}$Anton Pannekoek Institute for Astronomy, University of Amsterdam, Science Park 904, 1098 XH, Amsterdam, The Netherlands\\
$^{10}$Department of Physics, Texas Tech University, Box 41051, Lubbock, TX 79409-1051, USA\\
$^{11}$Leiden Observatory, Leiden University, Neils Bohrweg 2, 2333 CA, Leiden, The Netherlands\\
$^{12}$ASTRON, the Netherlands Institute for Radio Astronomy, Postbus 2, 7900 AA, Dwingeloo, the Netherlands\\
$^{13}$Dipartimento di Fisica, Universit\`a degli Studi di Cagliari, SP Monserrato-Sestu km 0.7, 09042 Monserrato, Italy
}
\date{Accepted XXX. Received YYY; in original form ZZZ}

\pubyear{2016}

\begin{document}
\label{firstpage}
\pagerange{\pageref{firstpage}--\pageref{lastpage}}
\maketitle

\begin{abstract}
We present the results of VLA, ATCA, and Swift XRT observations of the 2015 outburst of the transient neutron star X-ray binary (NSXB), EXO 1745$-$248, located in the globular cluster Terzan 5. Combining (near-) simultaneous radio and X-ray measurements we measure a correlation between the radio and X-ray luminosities of $L_R\propto L_X^\beta$ with $\beta=1.68^{+0.10}_{-0.09}$, linking the accretion flow (probed by X-ray luminosity) and the compact jet (probed by radio luminosity). While such a relationship has been studied in multiple black hole X-ray binaries (BHXBs), this work marks only the third NSXB with such a measurement. Constraints on this relationship in NSXBs are strongly needed, as comparing this correlation between different classes of XB systems is key in understanding the properties that affect the jet production process in accreting objects.
Our best fit disc-jet coupling index for EXO 1745$-$248 is consistent with the measured correlation in NSXB 4U 1728$-$34 ($\beta=1.5\pm 0.2$) but inconsistent with the correlation we fit using the most recent measurements from the literature of NSXB Aql X-1 ($\beta=0.76^{+0.14}_{-0.15}$). While a similar disc-jet coupling index appears to hold across multiple BHXBs in the hard accretion state, this does not appear to be the case with the three NSXBs measured so far. Additionally, the normalization of the EXO 1745$-$248 correlation is lower than the other two NSXBs, making it one of the most radio faint XBs ever detected in the hard state.
We also report the detection of a type-I X-ray burst during this outburst, where the decay timescale is consistent with hydrogen burning.
\end{abstract}



\section{Introduction}
The accretion process onto compact objects and the production of relativistic jets are fundamentally connected.
Low mass X-ray binaries (XBs), which contain a stellar-mass compact object, such as a black hole (BH) or neutron star (NS), accreting from a companion star,
are ideal candidates to study this relationship, as the rapid (day--week) outburst timescales of these systems allow us to track accretion and jet behaviour in real time.

Multi-wavelength studies of XBs have linked changes in the accretion flow (probed by spectral and variability properties of the X-ray emission) to those in the jet (probed by radio emission, e.g., \citealt{migfen06,tudose09,millerj12,corb13}).
In BHXB systems, a phenomenological model has been put forward to explain this connection, where changes in mass accretion rate are the catalyst driving changes in jet behaviour \citep{tan72,blandford79,vadawale2003,fenbelgal04,fenhombel09}. In the hard X-ray accretion state an optically-thick, steady, compact jet is present.  As the mass accretion rate increases during the rise of the outburst, the jet velocity and power are also thought to increase (although this has not yet been directly proven from observational data, as measurements of jet velocity in the hard state are difficult to make). When the source makes the transition from hard to soft accretion states at higher luminosities
, the system launches discrete, optically-thin, relativistically-moving ejecta, possibly as a result of internal shocks in the jet flow produced by the changes in jet velocity. The compact jet is quenched as the source moves into the softer accretion state, and then re-established as the source moves back into the hard state (where the jet is re-established well before quiescence; \citealt{kalem13}).

Similar to BHXBs, NSXB outburst behaviour is also thought to be governed by mass accretion rate (e.g., \citealt{homan10}). In terms of the connection between inflow and outflow, NSXB and BHXB systems display both similarities and differences \citep{migfen06}. A steady, compact jet is observed in NSXBs at lower X-ray luminosities ($<0.1\, L_{\rm edd}$) in hard accretion states (i.e., island accretion states; \citealt{mig10}), and discrete jet ejections have been found at higher X-ray luminosities (typically seen in Z sources persistently accreting at high fractions of Eddington; e.g., \citealt{fenwu04}, \citealt{foma01,spen13}), as seen in BHXBs. However, BHXBs tend to be much more radio loud than NSXBs at the same X-ray luminosity (\citealt{fenkuul01}; \citealt{migfen06}). While this could imply NS jets are less powerful, \cite{kord2006} suggest that jet power is comparable in NS/BH systems, and properties such as the mass of the compact object or radiative efficiency are responsible for the different radio luminosity levels. Additionally, NSXB jets do not all appear to be fully quenched in softer accretion states as they are in BHXBs \citep{mig04}. While commonalities could indicate that the physical mechanism (possibly related to the mass accretion rate) powering the jets in both classes of system is similar, differences suggest the nature of the compact object still may play an important role. Analyzing and quantifying the similarities and differences between these systems is key to understanding the properties that affect the jet production process (e.g., mass, spin, existence of a surface or event horizon) across all scales.

A key observational tracer of the accretion-jet connection in XBs is the correlation found between radio and X-ray luminosities in the hard state ($L_R\propto L_X^\beta$, where $\beta$ represents the disc-jet coupling index, e.g., $\beta_{\rm BH}\sim0.6$; \citealt{cor03}; \citealt{galfenpol03}; \citealt{corb13})\footnote{We note that \citealt{corr11h} present evidence for two different tracks in this correlation for BH systems, a radio-loud and radio-quiet track, although, recent work by \citealt{gallo14} found that a two track description is only statistically preferred when luminosity errors are $<0.3$ dex.}.
This non-linear correlation is consistent with scale-invariant jet models, where a self-absorbed synchrotron jet is coupled to an %
accretion flow, total jet power is a fixed fraction of the accretion power, and X-ray luminosity depends on mass accretion rate (\citealt{falke95}; \citealt{heisun03}; \citealt{mar03aa}).  Further, through the addition of a mass parameter, this correlation has been extended across the mass scale to include AGN, the supermassive analogues of BHXBs \citep{merhezdi03}; $\log(L_X)= \xi_R \log(\nu L_R)- \xi_M \log M_{BH}+B$, where the coefficients, $\xi_R=1.45\pm0.04$, $\xi_M=0.88\pm0.06$, and $B=-6.07\pm1.10$ (\citealt{falk4}; \citealt{plot12}). 

The $L_R\propto L_X^\beta$ correlation has been shown to hold in multiple BHXBs from quiescent luminosities as low as $10^{-9}\, L_{\rm edd}$ to outburst luminosities as high as $10^{-2}\, L_{\rm edd}$, above which the compact jet is quenched (we note that while the correlation holds tightly in individual systems, there is more scatter when the whole sample of BHXBs is considered together; \citealt{gallo14,plotgal15}). However, our knowledge of this correlation in individual NS systems is limited. Two NSXBs (4U 1728$-$34 and Aql X$-$1), have measured correlations, including data spanning only one order of magnitude in X-ray luminosity \citep{migfen06}. While 4U 1728$-$34 shows a correlation of $L_R\propto L_X^{1.5}$ \citep{mig03}, consistent with what we would expect from radiatively efficient accretion due to the NS's surface, there have been conflicting results for this correlation in Aql X-1. \citet{tudose09} measured $L_R\propto L_X^{0.4}$ for Aql X-1, which is more consistent with radiatively inefficient accretion flows (like those seen in BHXBs). However, \citet{tudose09} took Aql X-1 data from a mixture of accretion states; \cite{migfen06} show the correlation is consistent with $L_R\propto L_X^{1.4}$ when including only data taken in the hard accretion states for both Aql X-1 and 4U 1728$-$34 (we note that the \cite{migfen06} correlation only included 2 data points from Aql X-1, while the data from the full hard state coverage of the outburst, presented in \cite{millerj10}, is more consistent with a flatter correlation). Including data from softer accretion states could account for the differing disc-jet coupling indices between Aql X$-$1 and 4U 1728$-$34, although we direct the reader to the discussion section of this paper for an updated correlation for Aql X-1 and discussion of this discrepancy.
Further, three transitional milli-second pulsars (tMSPs;  binary NS systems that have been found to switch from a rotation powered pulsar state to an accreting XB state), have recently been shown to all lie on a shallower correlation, $L_R\propto L_X^{0.7}$, distinct from hard state NSXBs and much more consistent with BHXBs \citep{deller2015}.   . 
In addition to the disc-jet coupling index, the intrinsic normalization of this correlation clearly varies between BHXBs, NSXBs and tMSPs as groups, and between individual BHXB systems \citep{gallo14}.
More well measured correlations, including normalization and disc-jet coupling indices are needed to determine which NSXB behaviour is the norm, and determine the mechanisms driving the difference between the correlations of hard state NSXBs and tMSPs. Here we report on the third individual NSXB radio/X-ray correlation measured to date, from data taken during the 2015 outburst of the NSXB EXO 1745-248, located in the globular cluster Terzan 5.

 \renewcommand\tabcolsep{3.pt}
\begin{table*}
\small
\caption{Summary of Swift XRT Observations and Fluxes of EXO 1745$-$248}\quad
\centering
\begin{tabular}{ ccccccccccccc }
 \hline\hline
  {\bf Obs ID}&{\bf Exp.}&{\bf Mode}&{\bf Date} &{\bf MJD}$^a$&{\bf Count Rate}&{\bf $\mathbf{N_H}$}&{\bf $\mathbf{{\bm\Gamma}}$}$^b$&{\bf kT}$^c$&{ \bf $\mathbf{F_{1-10 {\rm \bf }}}$$^{d,e}$}&{\bf $\mathbf{\chi_{\nu}^2}$}&{\bf dof}\\
   &{\bf Time (s)}&&{\bf (2015)}&{\bf}&{\bf($\mathbf{{\rm \bf cnt\,s}^{-1}}$)}&{\bf $\mathbf{(10^{22}\,{\rm\bf  cm}^{-2})}$}&{\bf}&{\bf (keV)}&{\bf ($\mathbf{10^{-10}{\rm \bf erg\,s}^{-1}{\rm \bf cm}^{-2}}$)}&&\\[0.2cm]
  \hline \underline{ {\bf Hard State}}\\[0.15cm]
\phantom{0}32148017$^f$&1965.36&PC&Mar 17&57098.7491&\phantom{0}\phantom{0}$0.73^{+0.03}_{-0.02}$&$4.41^{+0.85}_{-0.74}$&$1.12^{+0.22}_{-0.21}$&\dots&\phantom{0}\phantom{0}$3.77^{+0.26}_{-0.23}$&1.01&\phantom{0}44\\[0.15cm]
32148021&1188.47&WT&Mar 21&57102.8797&\phantom{0}\phantom{0}$9.11^{+0.19}_{-0.22}$&$3.36^{+0.20}_{-0.19}$&$1.24^{+0.08}_{-0.07}$&\dots&\phantom{0}$14.19^{+0.32}_{-0.32}$&1.09&117\\[0.15cm]
32148023&\phantom{0}771.57&WT&Mar 23&57104.0629&\phantom{0}\phantom{0}$7.33^{+0.26}_{-0.29}$&$3.01^{+0.29}_{-0.27}$&$1.31^{+0.11}_{-0.10}$&\dots&\phantom{0}$10.61^{+0.36}_{-0.34}$&1.07&\phantom{0}62\\[0.15cm]
\phantom{0}32148024$^g$&\phantom{0}898.50&WT&Mar 25&57106.2081&\phantom{0}\phantom{0}$9.12^{+0.20}_{-0.22}$&$3.11^{+0.22}_{-0.21}$&$1.22^{+0.08}_{-0.08}$&\dots&\phantom{0}$13.73^{+0.36}_{-0.35}$&1.09&\phantom{0}90\\[0.15cm]
32148024&\phantom{0}323.36&WT&Mar 25&57106.2564&\phantom{0}\phantom{0}$8.66^{+0.25}_{-0.27}$&$2.65^{+0.37}_{-0.33}$&$1.13^{+0.15}_{-0.14}$&\dots&\phantom{0}$12.49^{+0.56}_{-0.55}$&0.87&\phantom{0}30\\[0.15cm]
32148025&\phantom{0}513.67&WT&Mar 26&57107.8624&\phantom{0}\phantom{0}$7.84^{+0.24}_{-0.26}$&$3.17^{+0.38}_{-0.35}$&$1.20^{+0.14}_{-0.14}$&\dots&\phantom{0}$11.84^{+0.45}_{-0.44}$&1.00&\phantom{0}44\\[0.15cm]
32148025&\phantom{0}534.59&WT&Mar 26&57107.9322&\phantom{0}\phantom{0}$7.21^{+0.24}_{-0.27}$&$2.85^{+0.33}_{-0.31}$&$1.16^{+0.13}_{-0.13}$&\dots&\phantom{0}$11.48^{+0.43}_{-0.43}$&1.01&\phantom{0}42\\[0.15cm]
32148026&\phantom{0}302.78&WT&Mar 28&57109.5274&\phantom{0}\phantom{0}$7.86^{+0.25}_{-0.26}$&$2.44^{+0.41}_{-0.37}$&$0.94^{+0.16}_{-0.15}$&\dots&\phantom{0}$11.79^{+0.56}_{-0.55}$&1.29&\phantom{0}42\\[0.15cm]
32148027&\phantom{0}344.07&WT&Apr 02&57114.4467&\phantom{0}\phantom{0}$7.76^{+0.29}_{-0.33}$&$2.43^{+0.38}_{-0.34}$&$1.00^{+0.15}_{-0.15}$&\dots&\phantom{0}$11.15^{+0.52}_{-0.51}$&0.95&\phantom{0}46\\[0.15cm]
32148029&\phantom{0}764.11&WT&Apr 06&57118.1135&\phantom{0}\phantom{0}$9.47^{+0.25}_{-0.27}$&$2.09^{+0.20}_{-0.18}$&$0.87^{+0.08}_{-0.08}$&\dots&\phantom{0}$13.74^{+0.37}_{-0.38}$&1.21&\phantom{0}80\\[0.15cm]
32148030&\phantom{0}443.32&WT&Apr 12&57124.3745&\phantom{0}$25.84^{+0.57}_{-0.63}$&$2.13^{+0.15}_{-0.14}$&$0.86^{+0.07}_{-0.07}$&\dots&\phantom{0}$38.96^{+0.87}_{-0.86}$&1.08&\phantom{0}99\\[0.15cm]
32148031&\phantom{0}704.54&WT&Apr 13&57125.8931&\phantom{0}$12.18^{+0.60}_{-0.66}$&$2.20^{+0.18}_{-0.18}$&$0.90^{+0.08}_{-0.08}$&\dots&\phantom{0}$29.05^{+0.74}_{-0.73}$&1.04&\phantom{0}98\\[0.15cm]
\phantom{0}32148032$^f$&\phantom{0}849.08&PC&Apr 14&57126.4853&\phantom{0}\phantom{0}$6.34^{+0.10}_{-0.09}$&$2.07^{+0.21}_{-0.20}$&$0.84^{+0.09}_{-0.09}$&\dots&\phantom{0}$43.41^{+1.33}_{-1.31}$&1.11&\phantom{0}48\\[0.15cm]
\hline \underline {\bf Soft State$^h$}\\[0.15cm]
32148033&\phantom{0}615.82&WT&Apr 20&57132.2902&\phantom{0}$95.13^{+1.63}_{-1.74}$&$2.47^{+0.05}_{-0.05}$&{ \dots}&${2.62^{+0.07}_{-0.06}}$&$121.39^{+1.19}_{-1.19}$&1.26&100\\[0.15cm]
32148033&\phantom{0}579.58&WT&Apr 20&57132.6898&$110.15^{+1.01}_{-1.09}$&$3.37^{+0.09}_{-0.09}$&${1.53^{+0.04}_{-0.04}}$&{ \dots}&$158.37^{+1.96}_{-1.91}$&1.38&107\\[0.15cm]
32148034&\phantom{0}385.78&WT&Apr 22&57134.1423&$106.36^{+1.15}_{-1.20}$&$2.44^{+0.06}_{-0.06}$&{ \dots}&${3.76^{+0.18}_{-0.16}}$&$140.70^{+1.76}_{-1.75}$&1.35&\phantom{0}45\\[0.15cm]
32148034&\phantom{0}290.99&WT&Apr 22&57134.2161&$113.53^{+1.26}_{-1.39}$&$2.95^{+0.15}_{-0.14}$&${0.93^{+0.06}_{-0.06}}$&{ \dots}&$184.63^{+3.01}_{-2.96}$&0.93&\phantom{0}36\\[0.15cm]
32148035&\phantom{0}376.85&WT&Apr 24&57136.4053&\phantom{0}$80.21^{+1.10}_{-1.17}$&${3.37^{+0.17}_{-0.16}}$&${1.53^{+0.07}_{-0.07}}$&{ \dots}&$112.37^{+2.27}_{-2.18}$&1.32&\phantom{0}31\\[0.15cm]
32148036&\phantom{0}988.60&WT&Apr 26&57138.1412&\phantom{0}$66.57^{+1.12}_{-1.20}$&${2.52^{+0.05}_{-0.05}}$&{ \dots}&${3.13^{+0.09}_{-0.09}}$&\phantom{0}\phantom{0}${81.7^{+0.08}_{-0.08}}$&{1.41}&\phantom{0}71\\[0.15cm]
\hline \underline {\bf Hard State}\\[0.15cm]
32148053&1106.30&WT&Jun 21&57194.6817&\phantom{0}\phantom{0}$0.14^{+0.03}_{-0.04}$&$1.66^{+1.17}_{-0.84}$&$1.44^{+0.74}_{-0.62}$&\dots&\phantom{0}\phantom{0}$0.32^{+0.06}_{-0.05}$&1.19&\phantom{0}\phantom{0}9\\[0.15cm]
\phantom{0}32148054$^i$&\phantom{0}\phantom{0}79.91&WT&Jun 24&57197.8643&\phantom{0}\phantom{0}$0.09^{+0.06}_{-0.03}$&$\dots$&$\dots$&\dots&\phantom{0}\phantom{0}$0.09^{+0.11}_{-0.06}$&\dots&\dots\\[0.2cm] \hline
\end{tabular}\\
\begin{flushleft}

{$^a$ All MJD values quoted represent the mid point of the observations.}\\
{$^b$ $\Gamma$ represents the power-law photon index. }\\
{$^c$ T represents the DISKBB temperature.}\\
{$^d$ Uncertainties are quoted at the $1\sigma$ level.}\\
{$^e$1--10 keV flux; see footnote 11 for discussion of why this X-ray band is chosen.}\\
{$^f$Note that this observation was piled-up; please see \S 2.1 for details.}\\
{$^g$Note that this observation contained an X-ray burst; please see \S 2.1 \& 4.2 for details. We excluded the burst interval when performing our spectral analysis of this observation.}\\
{$^h$Note that soft state measurements are not included in our radio/X-ray correlation analysis. We show them here for comparison purposes and to clearly show the transition between the hard and soft state.}\\
{$^i$Due to the limited exposure time of this observation, the flux for this observation is determined by using the model fits from the previous observation.}\\
\end{flushleft}
\label{table:imfluxx}
\end{table*}
\renewcommand\tabcolsep{6.0pt}

 \subsection{Terzan 5: EXO 1745-248}
 Terzan 5 is a massive ($\sim 10^6 M_\odot$; \citealt{lanz10}) globular cluster located in the Galactic centre region (distance of $5.9\pm0.5$ kpc; \citealt{val07}), with a high stellar density, leading to a very high stellar encounter rate (the highest measured so far; \citealt{bah13}). This cluster contains three transient X-ray sources confirmed to be accreting NSs, EXO 1745$-$248 (Terzan 5 X-1), IGR J17480$-$2446 (Terzan 5 X-2), and Swift J174805.3$-$244637 (Terzan 5 X-3; \citealt{wij05,bord10,st10,degw12,bah14}), as well as several other detected quiescent X-ray sources \citep{heinke06}. Historically, X-ray activity was first detected from Terzan 5 in 1980, in the form of multiple X-ray bursts, indicating the presence of an outbursting NSXB \citep{mak81,in84}. Subsequent X-ray activity was observed in 1984, 1990, 1991, 2000, 2002, 2010, 2011, and 2012, where activity in 2000 and 2011\footnote{The 2011 outburst showed superburst activity \citep{alta12}.}  was attributed to EXO 1745$-$248\footnote{We note that it is not known whether the Terzan 5 outbursts in the early 80s and 90s are associated with EXO 1745$-$248.} (\citealt{heinke03,alt12,ser12}; see Table 1 in \citealt{degw12} and references therein for a review of past X-ray activity in Terzan 5).

\begin{table*}
\small
\caption{Summary of Radio Frequency Observations and Flux Densities of EXO 1745$-$248}\quad

\begin{tabular}{ ccccccc }
 \hline\hline
  {\bf Telescope}&{\bf Date} &{\bf MJD$^a$}&{\bf Freq.}&{ \bf Flux$^{b,c}$}&{\bf Spectral}\\
   &{\bf (2015)}&&{\bf (GHz)}&{\bf($\mathbf{\bm \mu {\rm \bf Jy}\,{\rm \bf bm}^{-1}}$)}&{\bf Index$^d$} \\[0.15cm]
  \hline
VLA&Mar 19&57100.43155&\phantom{0}9.0&\phantom{0}28.7$\pm$6.0&\dots\\[0.1cm]
VLA&Mar 19&57100.43155&11.0&\phantom{0}22.8$\pm$8.0&-1.20$\pm$1.97\\[0.1cm]
VLA&Mar 24&57105.53915&\phantom{0}9.0&\phantom{0}47.8$\pm$6.0&\dots\\[0.1cm]
VLA&Mar 24&57105.53915&11.0&\phantom{0}30.3$\pm$8.0&-2.20$\pm$1.4\\[0.1cm]
VLA&Apr 12&57124.40413&\phantom{0}9.0&238.1$\pm$8.3&\dots\\[0.1cm] 
VLA&Apr 12&57124.40413&11.0&247.6$\pm$9.3&0.15$\pm$0.26\\[0.1cm] 
ATCA&Apr 16&57128.75694&\phantom{0}5.5&372.0$\pm$7.0&\dots\\[0.1cm]
ATCA&Apr 16&57128.75694&\phantom{0}9.0&340.0$\pm$7.8&-0.18$\pm$0.06\\[0.1cm]
ATCA&Jun 23&\phantom{0}57196.60938$^e$&\phantom{0}5.5&$<17$&\dots\\[0.1cm]
ATCA&Jun 23&\phantom{0}57196.60938$^e$&\phantom{0}9.0&$<19$&\dots\\[0.15cm]\hline
\end{tabular}\\
\begin{flushleft}
{$^a$ All MJD values quoted represent the mid point of the observations.}\\
{$^b$ Uncertainties are quoted at the $1\sigma$ level.}\\
{$^c$ Radio flux density, where uncertainties quoted include the 1\% systematic errors appropriate to both VLA X-band observations and the ATCA  3/6 cm observations.}\\
{$^d$ All spectral indices given use the formalism, $f_\nu\propto\nu^\alpha$; where $\alpha$ is the spectral index.}\\
{$^e$ The source was not detected in these observations, fluxes presented here are $3\sigma$ upper limits.}\\
\end{flushleft}

\label{table:imfluxr}
\end{table*}

 On 2015 March 13, renewed X-ray activity from Terzan 5 was detected \citep{alt15} by the Swift Burst Alert Telescope (BAT; \citealt{kr13}) transient monitor. While the X-ray position from follow up Swift X-ray Telescope (XRT) observations \citep{bah15}  
was consistent with EXO 1745$-$248, IGR J17480$-$2446, and several other quiescent X-ray sources \citep{heinke06}, the spectrum showed a higher than typical hydrogen column density, $N_H=4\pm0.8\times10^{22}$, for sources in Terzan 5 (\citealt{bah14}), consistent with previous observations of EXO 1745$-$248 \citep{kuulk03}.  \citet{lin15} measured a refined Swift XRT source position centered on the known X-ray position of EXO 1745$-$248 ($2\farcs2$ error circle) further suggesting that the outbursting source in Terzan 5 was in fact EXO 1745$-$248.  \citet{trem15} detected a radio counterpart with observations by the Karl G. Jansky Very Large Array (VLA). These radio observations, which localized the source within $0\farcs4$ of the published Chandra coordinates (source CX3 in \citealt{heinke06}), and later optical observations that identified the optical counterpart during this outburst \citep{ferr15}, confirmed the identification by \citet{lin15}. 

We obtained multiple epochs of (near-) simultaneous VLA, Australia Telescope Compact Array (ATCA), and Swift XRT observations during the 2015 outburst of EXO 1745$-$248. 
 In \S 2 we describe the data collection and reduction processes. In \S 3 we present a refined radio position of EXO 1745$-$248, measurements of the jet spectral index, and the radio/X-ray correlation in this source. \S 4 contains an interpretation of this correlation, comparison to other NS and BH XB sources,  and an analysis of an X-ray burst detected in one of the Swift XRT observations. A summary of the results is presented in \S 5.

\section{Observations and Data Analysis}
\subsection{X-ray Observations}

\begin{figure*}
\centering
{\includegraphics[width=19cm,height=10cm]{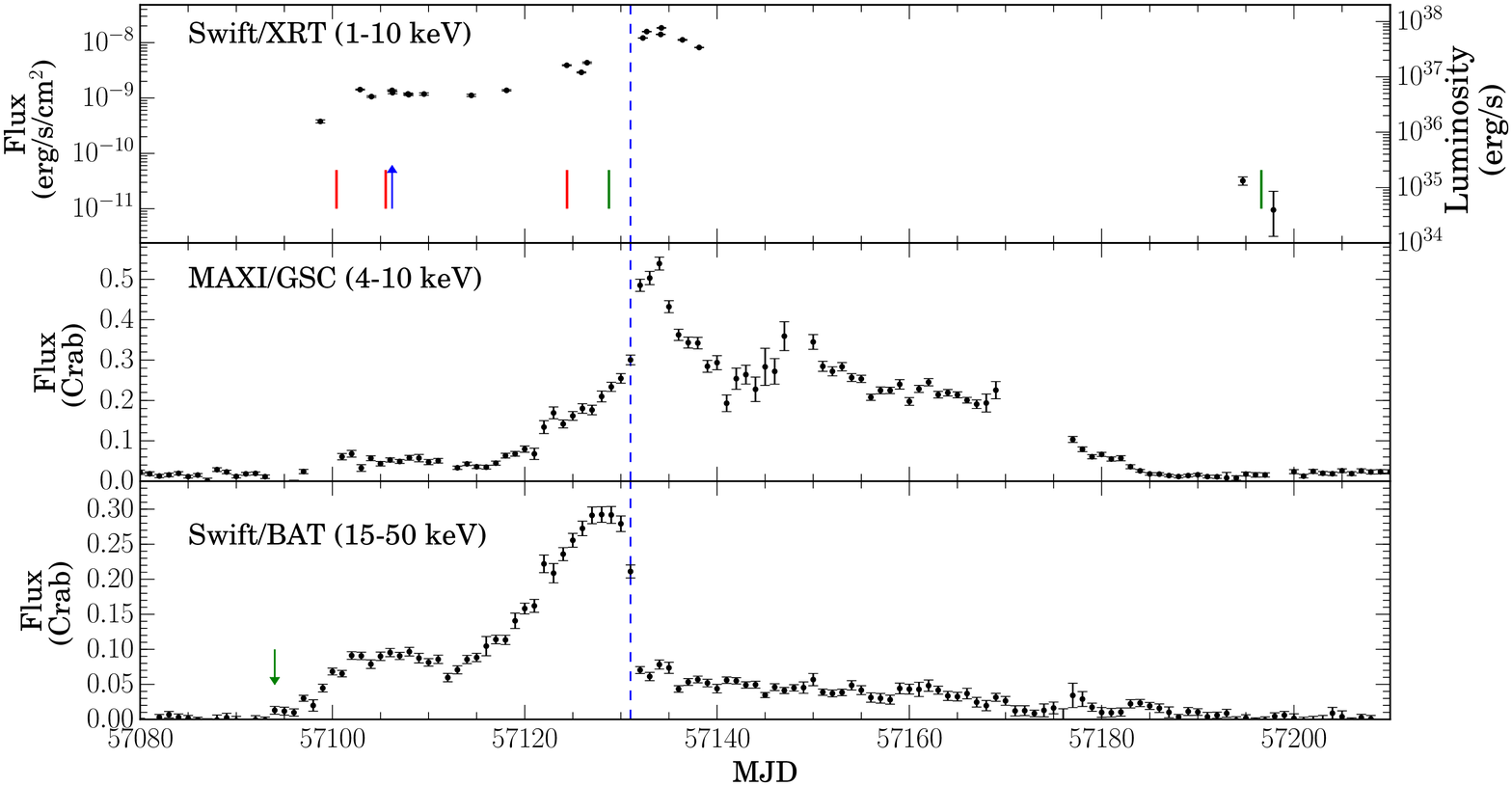}}
\caption{\label{fig:xlight} X-ray light curves of the 2015 outburst from EXO 1745$-$248 as seen by Swift/XRT (top), MAXI/GSC (middle; http://maxi.riken.jp) and Swift/BAT (bottom; \citealt{kr13}). Swift/XRT fluxes are derived from spectral fitting (see \S 2.1 in text). To calculate luminosity in the top panel, we assumed a distance of 5.9 kpc. The blue dashed line shows the apparent hard to soft state-transition at MJD 57131. Vertical bars in the top panel indicate the time of radio observations by the ATCA (green) and VLA (red). The blue arrow in the top panel indicates the time of the detected X-ray burst, and the green arrow in the bottom panel indicates indicates the time of the first detection of the outburst in Swift BAT. Note that while we only use the hard state measurements in our radio/X-ray correlation analysis, we show some of the soft state measurements near the times of our radio observations to clearly show the hard-soft transition and the luminosity at that transition.}
\end{figure*}

We monitored the outburst of EXO 1745$-$248 multiple times per week with Swift XRT following its detection in 2015. This paper considers only those observations that are most relevant to analyzing the radio/X-ray correlation and accretion state transition. We summarize these observations in Table~\ref{table:imfluxx} and Figure~\ref{fig:xlight}; these consist of two observations in photon counting (PC) mode, which produces 2D images, and 19 observations in windowed timing (WT) mode, which collapses data to 1-dimension for fast readout.

We used HEASOFT 16.6 and FTOOLS \footnote{http://heasarc.gsfc.nasa.gov/ftools/} \citep{black95} for all data reduction and analysis. All Swift XRT observations were reprocessed via \texttt{xrtpipeline} and \texttt{xselect} was used to manually extract source and background spectra. We used \texttt{xrtmkarf} to produce ancillary response files. Finally, we performed spectral analysis using XSPEC 12.8.2 \citep{ar96} in the 0.3--10 keV band for PC mode data and the 0.6--10 keV band for WT mode data.

Our PC mode observations in this campaign were heavily piled-up due to the high count rate of the source. Thus we followed the UK Swift Science Data Centre pile-up thread\footnote{http://www.swift.ac.uk/analysis/xrt/pileup.php} and extracted source spectra from an annulus (13--70" for the first PC mode observation and 20--100" for the second PC observation), excluding the piled-up region in these observations. The PC mode observations only showed evidence for one bright source.

For heavily absorbed sources, WT data show low energy spectral residuals, which can cause spectral uncertainties in the $ \lesssim 1.0$ keV region\footnote{http://www.swift.ac.uk/analysis/xrt/digest\_cal.php\#abs}. These residuals mostly affect grade 1 events and above, and events below $\sim$ 0.6 keV. Thus for our WT mode data, we extracted spectra only from grade 0 events and excluded events below 0.6 keV.

We extracted a spectrum from each observation separately and performed spectral fitting. Our main model for spectral analysis is an absorbed powerlaw (TBABS*PEGPOWERLAW in XSPEC), where we assume the cross sections from \citet{vern96} and abundances from \citet{wilms00}.
We use a comparison of Swift/BAT and MAXI light curves (Figure~\ref{fig:xlight}) to aid in defining which observations are in the hard/soft accretion state. This comparison reveals a large drop in the hard flux simultaneous with a rise in the soft flux, indicative of a hard-to-soft state transition on MJD 57131. Thus for spectra from Swift/XRT observations after this point, we tried both absorbed powerlaw and absorbed disc blackbody (TBABS*DISKBB in XSPEC) models, and chose the fit with lower $\chi^2$ for this study. Although a two-component model (e.g., DISKBB$+$PEGPOWLAW) is often used to fit NSXB soft states, we are only interested in identifying the dominant component (as opposed to performing a detailed characterization of the spectrum) and obtaining a flux estimate. Thus we only fit simple one component models for the purpose of this work.

We note that the MAXI and Swift/BAT data do not clearly show the soft-to-hard state transition, probably because it occurred at a lower luminosity where the S/N of these instruments is low.  However, all observed XB outbursts return to the hard state at luminosities above $10^{35}\, {\rm erg\,s}^{-1}$ \citep{macar03,tetarenkob2015}, so we conclude it is extremely likely that the last two Swift data points, and the ATCA measurement between them, occurred during the hard state. The power law index measured for the June 21 observation, which was more consistent with the hard state observations than the soft state observations, support this conclusion.
Swift/XRT observations and results of our spectral analysis are reported in Table~\ref{table:imfluxx}.

We also detected an X-ray burst during the Mar 25 observation, which is discussed in detail in \S 4.2.

\subsection{Radio Observations}
\subsubsection{VLA}
Terzan 5 was observed with the VLA (Project Code: 14B-216) in three epochs, 2015 March 19, March 24, and April 12. The array was in the B configuration, with a resolution of $0\farcs6$, and we had 25.9 min on source for each epoch. All observations were made with the 3-bit samplers in X band ($8-12\,{\rm GHz}$), comprised of 2 base-bands, each with 16 spectral windows of 64 2-MHz channels each, giving a total bandwidth of 2.048 GHz per base-band. Flagging, calibration and imaging of the data were carried out within the Common Astronomy Software Application package (CASA\footnote{http://casa.nrao.edu}; \citealt{mc07}) using standard procedures. When imaging we used a natural weighting scheme to maximize sensitivity, two Taylor terms (nterms=2) to account for the large fractional bandwidth, and did not perform any self-calibration. We used 3C286 (J1331$+$305) as a flux calibrator and J1751$-$2524 as a phase calibrator. Flux densities of the source were measured by fitting a point source in the image plane (Stokes $I$ with the \textsc{imfit} task), and, as is standard for VLA X band data, systematic errors of 1\% were added. All flux density measurements are reported in Table~\ref{table:imfluxr}.

\subsubsection{ATCA}
During the 2015 outburst of EXO 1745$-$248, Terzan 5 was observed with the ATCA (Project Code: C2877) in two epochs, 2015 April 16 and June 23. The array was in the 6A configuration (resolution of $1\farcs89$/$1\farcs16$ arcsec at 5.5/9 GHz) in the first epoch, and the 6D configuration (resolution of $1\farcs91$/$1\farcs16$ arcsec at 5.5/9 GHz) in the second epoch. We had 8.0 hrs on source for both epochs. All observations were carried out at 5.5 and 9 GHz simultaneously, where each frequency band is comprised of 2048 1-MHz channels, giving a total bandwidth of 2.048 GHz per frequency band. Flagging and  calibration were carried out with the Multichannel Image Reconstruction, Image Analysis and Display (MIRIAD) software \citep{sault95}, using standard procedures. We used 1934-638 as a flux calibrator and 1748-253 as a phase calibrator. 
Imaging of the data was carried out within CASA using a Briggs weighting scheme (robust=1) and two Taylor terms (nterms=2). We did not perform any self-calibration. Flux densities of the source were measured by fitting a point source in the image plane (Stokes $I$ with the \textsc{imfit} task), and, as is standard for ATCA data, systematic errors of 1\% were added. All flux density measurements are reported in Table~\ref{table:imfluxr}.

\section{Results}
\subsection{Radio Source Position}
Through stacking all three epochs of our VLA data in the uv-plane, we measure a refined radio position of EXO 1745$-$248 to be the following (J2000),
\begin{eqnarray}\nonumber 
{\rm RA:}&\,17^{\rm h}48^{\rm m}05^{\rm s}.22467\pm 0.00084 \pm 0.01 \\ \nonumber
{\rm DEC:}&\,-24^{\circ}46'47\farcs666\pm 0.033 \pm 0.06\nonumber
\end{eqnarray}
where the quoted error bars represent the statistical error from fitting in the image plane and the nominal systematic uncertainties of $10\%$ of the beam size, respectively. 

This source position is within $0\farcs33$ of the published X-ray location of EXO 1745-248 (CX 3 in \citealt{heinke06}; RA/DEC errors 0.002s/0.02 arcsec), and within $0\farcs10$ of the optical location of EXO 1745-248 (\citealt{ferr15}; RA/DEC errors 0.01s/0.2 arcsec). The radio source is clearly unassociated with the two other previously identified NSXBs in Terzan 5; it is $2\farcs4$ away from IGR J17480$-$2446 (CX 24 in \citealt{heinke06}; RA/DEC errors 0.005s/0.09 arcsec) and $10\farcs3$ away from Swift J174805.3$-$244637 (\citealt{bah14}; RA/DEC errors 0.02s/0.2 arcsec).

\subsection{Jet Spectral Indices}
To obtain the jet spectral indices we fit (linearly in log space) a power-law to the derived radio flux densities (between the two-base-bands in the VLA data and between 5.5 and 9 GHz in the ATCA data) against frequency at each epoch ($f_\nu \propto \nu^\alpha$; where $\alpha$ is the spectral index). 
All spectral index measurements are reported in Table~\ref{table:imfluxr}. In the March 19 and April 12 VLA epochs, the spectral index measurements are consistent with a flat ($\alpha=0$) or slightly inverted ($\alpha>0$) spectrum, although, in the March 24 VLA epoch and the ATCA epoch on Apr 16, the spectral index appears to be more consistent with a slightly steeper index ($\alpha<0$). 
However, both the March 19 and March 24 VLA epochs have large uncertainties (due to the low signal-to-noise ratio and small lever arm in frequency) that make it impossible to conclusively distinguish between steep, flat, or an inverted spectra.   
A flat or slightly inverted spectrum, commonly seen from compact jets during hard accretion states in BHXBs \citep{fender01} and some NSXBs (e.g., \citealt{migfen06,mig10}), is believed to be produced as the result of the superposition of many different synchrotron components originating from different regions along the jet (e.g., \citealt{blandford79}).

 \begin{figure}
\centering
 {\includegraphics[width=9cm,height=7cm]{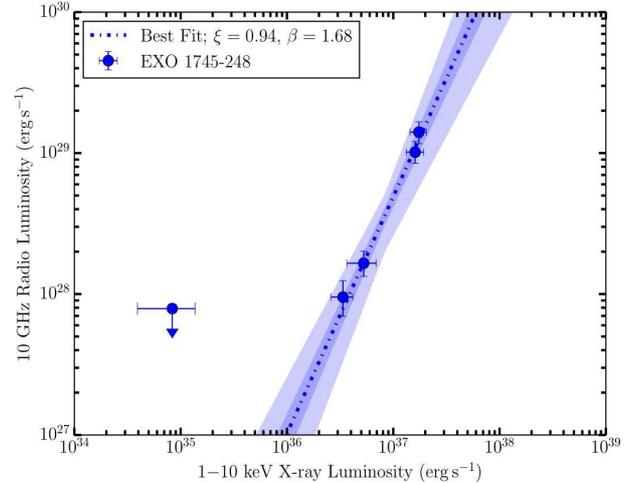}}
\caption{\label{fig:fitlrlx} Radio/X-ray correlation during the hard accretion state of the 2015 outburst of EXO 1745$-$248. The dash-dotted line indicates the best fit using our MCMC techniques (see text for best fit parameters and uncertainties). The shaded regions represent the $1\sigma$ (dark blue) and $3\sigma$ (light blue) confidence intervals of the regression. Note that we do include the upper limit data point in our fit. The luminosities displayed here are calculated assuming a distance of 5.9 kpc.}
\end{figure}

\subsection{Radio X-Ray Correlation in EXO 1745$-$248}

\begin{table}
\small
\caption{Radio and interpolated X-ray fluxes of EXO 1745$-$248 used in the radio/X-ray correlation analysis}\quad
\centering
\begin{tabular}{ ccc }
 \hline\hline
 {\bf MJD}&{ \bf $\mathbf{F_{10 {\rm \bf GHz}}}$}$^{a,b}$&{ \bf $\mathbf{F_{1-10 {\rm \bf keV}}}$$^{a,c}$}\\
   &{\bf($\mathbf{\bm \mu {\rm \bf Jy}\,{\rm \bf bm}^{-1}}$)}& {\bf ($\mathbf{10^{-10}{\rm \bf erg\,s}^{-1}{\rm \bf cm}^{-2}}$)}\\[0.15cm]
  \hline
57100.43155&\phantom{0}23.2$\pm$5.0&\phantom{0}$8.02^{+0.41}_{-0.40}$\\[0.1cm]
57105.53915&\phantom{0}40.0$\pm$4.0&$12.78^{+0.67}_{-0.69}$\\[0.1cm]
57124.40413&245.3$\pm$5.6&$38.78^{+0.82}_{-0.89}$\\[0.1cm]
57128.75694&340.0$\pm$7.8&$41.89^{+1.39}_{-1.47}$\\[0.1cm]
\phantom{0}57196.60938$^d$&$<19$&\phantom{0}$0.20^{+0.13}_{-0.11}$\\[0.15cm]\hline
\end{tabular}\\
\begin{flushleft}
{$^a$ Uncertainties are quoted at the $1\sigma$ level.}\\
{$^b$ 10 GHz radio flux from combining the 2 base-bands.}\\
{$^c$ Interpolated X-ray fluxes in the 1--10 keV band.}\\
{$^d$ The source was not detected in this observation, the flux presented here is $3\sigma$ upper limit.}\\
\end{flushleft}

\label{table:fitflux}
\end{table}

To fit the radio/X-ray correlation in EXO 1745$-$248, we use radio and X-ray luminosities (spanning $\sim1$ dex in X-ray luminosity\footnote{While our radio observations span $\sim3$ dex in X-ray luminosity, our lowest luminosity point only has an upper limit on radio luminosity and thus is not very constraining.}) at 10 GHz (combined base-band measurements) and 1--10 keV, respectively\footnote{In previous studies that compute the radio/X-ray correlation, the X-ray energy band used can vary from author to author, but in this work we choose the 1--10 keV band, as we have found that this band is most commonly used in recent literature; e.g., \citealt{gallo14,corb13,deller2015}.}, and a Markov Chain Monte Carlo (MCMC\footnote{In this work, all of our codes use the {\tt emcee} python package to implement the MCMC algorithms \citep{hogg,for2013}.}) fitting algorithm. To properly account for uncertainties in both distance ($5.9\pm0.5$ kpc; \citealt{val07}) and flux, we build a hierarchical model within our MCMC, where we include distance as an additional parameter. This in turn allows us to calculate luminosities using our measured fluxes/uncertainties and samples drawn from the distance distribution (i.e., a Gaussian with mean of 5.9 and standard deviation of 0.5), and then perform a linear fit in log space on these luminosities.
Although many previous studies only compare the X-ray measurements closest in time to the radio measurements, our method takes a more conservative approach to data that is not strictly simultaneous.
In particular, as our XRT X-ray observations were not strictly simultaneous with the VLA radio observations, we use a MCMC linear interpolation method to estimate X-ray fluxes at the times of the radio observations. However, as the X-ray flux of outbursting NSXBs can vary on timescales of less than a day (the maximum separation between our radio and X-ray observations), our linear interpolation method may underestimate the uncertainties on the interpolated X-ray fluxes. Therefore, we conservatively scale the uncertainties on the interpolated X-ray fluxes to cover the full flux range of the neighbouring X-ray data (see Table~\ref{table:fitflux} for radio and interpolated X-ray fluxes used in our MCMC fitting).

We follow \cite{gallo14} when performing our MCMC fit with the following functional form,
\begin{equation}
(\log L_R -\log L_{R,c}) =\log\xi + \beta (\log L_X -\log L_{X,c})
\end{equation}
where, $L_R$ and $L_X$ are radio (10 GHz) and X-ray (1--10keV) luminosity, respectively, centering values $L_{R,c}=3.89\times10^{28}\, {\rm erg\,s}^{-1}$ and $L_{X,c}=8.38\times10^{36}\, {\rm erg\,s}^{-1}$ are the geometric means of the simultaneous radio and X-ray luminosity measurements (not including the upper limit data point), $\xi$ represents the normalization constant and $\beta$ represents the disc-jet coupling index. To include the upper limit data point in our fit, and better constrain the normalization and disc-jet coupling index, we add a condition in our log probability that does not allow solutions where, at the X-ray luminosity of the upper limit data point, the corresponding radio luminosity would exceed the upper limit value. Our best fit parameters 
are, normalization $\xi =0.94^{+0.14}_{-0.13}$ and disc-jet coupling index $\beta=1.68^{+0.10}_{-0.09}$, where uncertainties are quoted at the 15th and 85th percentiles (as done in \citealt{gallo14}; also see Figure~\ref{fig:fitlrlx}).

 \begin{figure*}
\centering
 {\includegraphics[width=16cm,height=12cm]{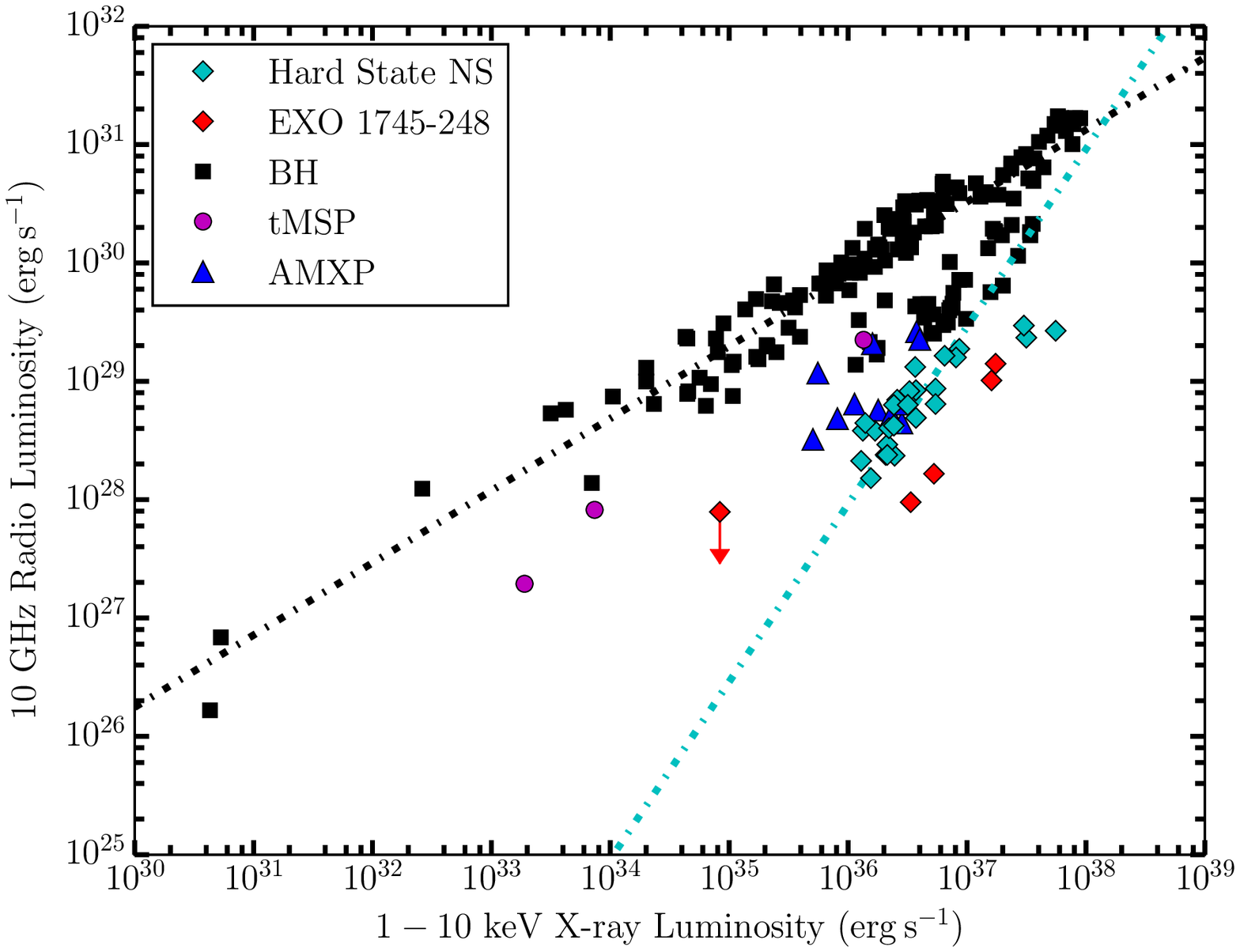}}
\caption{\label{fig:lrlx}Radio/X-ray correlation for different types of accreting stellar mass compact objects. Data points from the literature include, BHs (\citealt{millj11}; \citealt{galmilfen12}; \citealt{ratti2012}; \citealt{corb13}; \citealt{gallo14}), hard state neutron stars \citep{migfen06,millerj10}, transitional binary milli-second pulsars (tMSPs) and accreting milli-second X-ray pulsars (AMXPs) (\citealt{hill2011}; \citealt{papitto2013}; \citealt{deller2015}). Note that to convert between different radio bands we assume a flat radio spectral index. 
The dot-dashed lines show the best fit relations for BH (${ \beta=0.61}$, black; \citealt{gallo14}) and hard state NS systems (${ \beta=1.40}$, cyan; \citealt{migfen06}). The new measurements of EXO 1745$-$248 reported in this paper (highlighted in red; luminosities assume a distance of 5.9 kpc) are more radio quiet and/or X-ray loud when compared with the other hard state NS measurments. Note that error bars are not included in this plot for clarity.}  
\end{figure*}

\section{Discussion}
In the framework of scale-invariant jet models coupled to an accretion flow, X-ray luminosity scales with mass accretion rate ($L_X\propto \dot{M}^q$)\footnote{Although this is a standard assumption in many papers, we point out two caveats for NSXBs. First, this assumes that the bolometric correction (in the hard state) remains constant so that the X-ray luminosity measured over limited energies is representative of the bolometric luminosity. Second, there may be multiple mass accretion rates (e.g., that in the disc versus that in a radial inflow) contributing to the X-ray luminosity of a NSXB, and it is unclear which of these would impact jet production.}, total jet power is a fraction of the accretion power ($Q_{\rm jet}=f\dot{M}c^2$), and the jet luminosity scales with jet power, according to (\citealt{falke95}; \citealt{heisun03}; \citealt{mar03aa}) ,
\begin{equation}
L_\nu\propto Q_{\rm jet}^\eta
\end{equation}
Here, $\eta=\frac{2p-(p+6)\alpha+13}{2(p+4)}$ depends on the power-law index of the electron energy distribution ($p$), and the jet spectral index ($\alpha$).
When the jet is observed in the radio regime this in turn implies,
\begin{equation}
L_R\propto L_X^{{\eta}/{q}}
\end{equation}
where radiatively efficient flows display $q=1$, and radiatively inefficient flows display $q=2-3$.  

In the previous section we reported a disc-jet coupling index of $1.68^{+0.10}_{-0.09}$ for EXO 1745$-$248, which is consistent with a radiatively efficient accretion flow (possibly due to the neutron star surface; see \citealt{migfen06} and references therein for discussion) coupled to a steady, compact jet (i.e., values of $q=1$, $2\lesssim p\lesssim3$, $-0.7\lesssim \alpha \lesssim0.1$ will produce values of $1.4<\eta<2.0$ within the confidence interval we derived for EXO 1745$-$248).

\subsection{Comparison to other Neutron Star and Black Hole Systems}
While several BHXBs have measured disc-jet coupling indices (e.g., \citealt{gallo14} combine data from 24 different BHXB systems to yield a best-fit disc jet coupling index of $0.61\pm0.03$), to date there are only two individual NSXBs with previously measured disc-jet coupling indices, Aql X-1 and 4U 1728$-$34. \cite{mig03} report a disc-jet coupling index of $1.5\pm0.2$ in 4U 1728$-$34, while different works report conflicting correlations for Aql X-1. The Aql X-1 data used to fit the correlation in \cite{tudose09} originates from mixed accretion states. While \citet{migfen06} find that 4U 1728$-$34 and Aql X-1 are well fit together with a disc-jet coupling index of $1.40\pm0.23$, this fit only includes two data points from Aql X-1. More recently \citet{migliari2011} reported that Aql X-1 is fit by a disc-jet coupling index of $\sim 0.6$ (with no errors reported). 

Therefore, we combined the most recent hard state Aql X-1 data from the literature, including the two measurements from \cite{migfen06}, as well as measurements from \citet{millerj10}, but excluding data with radio upper limits or hard X-ray colour\footnote{Hard X-ray colour is defined in \citet{millerj10} as the count rate ratio between the 9.0-16.0 keV and 6.0-9.7 kev bands.} $<0.75$. We find a disc-jet coupling index of $0.76^{+0.14}_{-0.15}$. This new Aql X-1 result is not consistent with the 4U 1728$-$34 result, and suggests that the use of mixed accretion state measurements in \citet{tudose09} is not the sole cause of the flatter disc-jet coupling index. Instead the disc-jet coupling index of Aql X-1 is more consistent with those of BHXBs. However, this correlation in Aql X-1 is only measured over $\sim 0.8$ dex, and we note that \cite{corb13} observed temporary excursions from the typical radio/X-ray correlation in BHXB GX 339-4 when measured over $<2$ dex in X-ray luminosity.

Our measurement for EXO 1745$-$248 is much more consistent with 4U 1728$-$34, rather than Aql X-1 or the BHXBs (see Figure~\ref{fig:lrlx}), where the EXO 1745$-$248 and 4U 1728$-$34 indices are what is expected from the model presented above for a radiatively efficient accretion flow coupled to a compact jet. Interestingly, Aql X-1 has (only once) shown evidence of X-ray pulsations \citep{cas07}, suggesting that it may be more similar to the AMXPs or tMSPs. 

\cite{deller2015}, recently combined radio and X-ray measurements for three tMSPs to fit a correlation of $L_R\propto L_X^{0.7}$ over $\sim3$ dex in X-ray luminosity,  which occupies a region of the radio/X-ray plane distinct from all the hard state NSXBs, like EXO 1745$-$248 (see Figure~\ref{fig:lrlx}). Given that there is only one data point for this correlation in each individual tMSP, we are forced to only consider the correlation of this entire sample; although, given the correlation in BHXBs, we might expect the sample correlation to have a larger scatter than one might find in an individual source. The disc-jet coupling indices of tMSPs as a group are much more consistent with Aql X-1 than with EXO 1745$-$248 or 4U 1728$-$34.
\cite{deller2015} suggest that tMSPs are undergoing a propeller accretion mode, where the pressure of in-falling material is balanced by the magnetic field of the NS, and the NS's rotation accelerates the inner disc, in turn causing the majority of the material to be ejected in outflows as opposed to falling inward. This theory can explain the radiatively inefficient jet dominated states seen at lower accretion rates in tMSPs (i.e. the tMSP correlation, $L_R\propto L_X^{0.7}$), which display a similar disc-jet coupling index as those of BHXBs, just at fainter radio luminosities (the offset between BHXBs and tMSPs could be due to differing jet power, radiative efficiency, compact object mass, or jet launching mechanisms). However, it is unknown whether this jet dominated state occurs in all NSXBs or if entrance into this state is solely dependent on intrinsic NS characteristics such as magnetic field strength or spin period. In the current published NSXB sample (excluding tMSPs), { only one correlation measurement (i.e., our lowest luminosity point in EXO 1745$-$248) probes X-ray luminosities ${\lesssim10^{35}\, {\rm  erg\,s}^{-1}}$}. However, this measurement only has an upper limit on radio luminosity. While this data point appears not to be consistent with the tMSP correlation, we are unable to definitely determine whether this point lies on the extrapolation of our hard state NSXB best fit correlation at lower X-ray luminosities or perhaps, is part of an intermediate regime where the disc-jet coupling index flattens out during the transition between a steeper and flatter index (as seen in the multiple BHXBs, H1743$-$322; \citealt{corr11h}, XTE J1752$-$223; \citealt{ratti2012}, and MAXI J1659$-$152; \citealt{jonk12}).

From Figure~\ref{fig:lrlx} it is also clear that EXO 1745$-$248 has a lower normalization compared to the other hard state NSs, Aql X-1 and 4U 1728$-$34, by about a factor of 5 in radio luminosity at the same X-ray luminosity. Among transient XBs measured in the hard state at $L_X>10^{36}\,{\rm erg\,s}^{-1}$, EXO 1745$-$248 is the most radio faint source reported to date. This differing normalization may be analagous to what is seen in BH sources, where different individual sources appear to have different normalizations \citep{gallo14}. We note that while this difference could arise from having a well-known distance for EXO 1745$-$248 compared to more uncertain distances to Aql X-1 and 4U 1728$-$34, the distances to Aql X-1 and 4U 1728$-$34 would have to increase by a factor of three if this was a distance effect alone, which seems unlikely. On the other hand, a factor of 5 lower in radio luminosity at a given X-ray luminosity requires masses lower by a factor of about 10 if the sources follow the fundamental plane of BH accretion. Since NSs do not have such a large range of masses, mass alone can not explain the lower luminosity of EXO 1745$-$248, unless NSs and BHs follow very different fundamental planes of accretion. 
Further, \citet{migliari2011} found a possible relation between spin frequency and jet power, with faster spinning neutron stars being more radio luminous. Based on its X-ray burst properties (\S 4.2), we expect EXO 1745$-$248 to have a typical spin (200--600 Hz). However, \citet{migliari2011} did not include the recent results from tMSPs. At $L_X\sim 10^{36} {\rm \, erg \, s^{-1}}$, the tMSP M28I (254 Hz; \citealt{papitto2013}) has a significantly higher radio luminosity than Aql X-1 (550 Hz; \citealt{watt08}).  While this compares a tMSP to a NS, we take this as evidence that spin alone also cannot explain how radio loud a NSXB will be. Thus it seems likely that a combination of factors (e.g., mass, spin, inclination, magnetic field, radiative efficiency) may be required to produce a given radio luminosity.

This highlights the need for more radio/X-ray measurements of NSXBs, especially at the lower end of the luminosity spectrum, to answer these open questions. However, we note that obtaining such observations is very difficult, given that these NSXB sources decay very quickly (timescales on the order of a few days) through this desired luminosity range of  $10^{34}-10^{36}\, {\rm erg\,s}^{-1}$, necessitating intensive monitoring of these sources.

 \subsection{X-ray Burst Analysis}
\begin{figure}
\centering
 {\includegraphics[width=8.5cm,height=11cm]{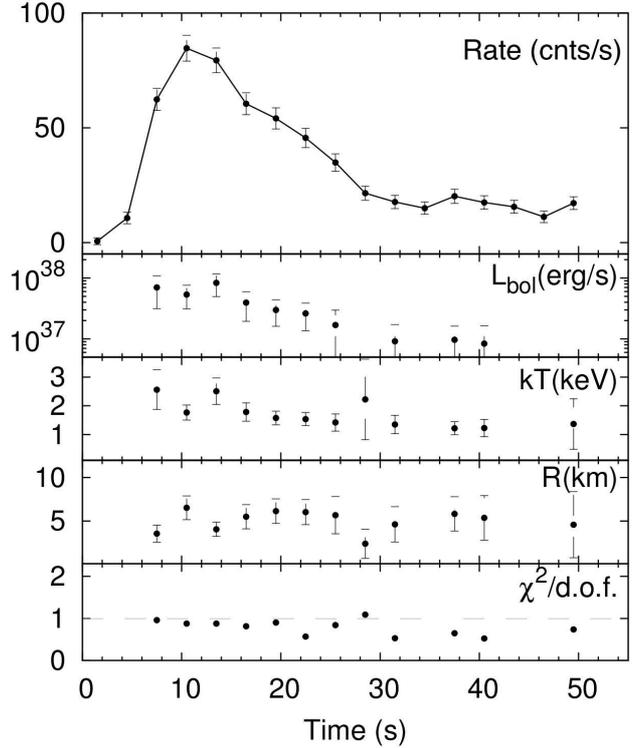}}
\caption{\label{fig:burst} Time-resolved spectroscopy of the detected type-I X-ray burst. An absorbed blackbody model was used to fit the data. We found evidence indicating slow cooling during the burst decay, however we found no evidence of photospheric radius expansion. Panels from top to bottom: Swift/XRT count rate, bolometric luminosity, temperature, apparent radius, reduced chi-squared of the spectral fit.  }
\end{figure}

 During our analysis we observed the presence of an X-ray burst, which we use here to further constrain the properties of this NSXB. Swift XRT detected an X-ray burst from Terzan 5 on 2015 Mar 25, with a net peak count rate at 04:56:42 UT of about $120\, {\rm cnts\, s}^{-1}$ (0.5-10 keV), on top of the persistent emission ($\sim10\, {\rm cnts \,s}^{-1}$). From the 0.5 s time-resolution light curve we estimate a rise time of 1.8 s (defined as the time to go from 25\% to 90\% of the net peak count rate). The burst lasted for about 25 s and then reached a ``plateau" for another $\sim$25 s, at a level higher than the pre-burst count rate. About 50 s after the burst onset the observation was interrupted (see Figure~\ref{fig:burst}).

To study the spectral evolution of the X-ray burst, we extracted a series of 3 s-long spectra from WT XRT data,
using a 100 s interval before the burst to subtract the persistent (source plus background) emission. We used a 20-pixel radius region to extract the
spectra, and verified that excluding the innermost 2 pixels (to correct for potential pile up) leads to consistent results.
We created an exposure map and ancilliary response file and used the latest response matrix from the calibration database. We grouped the resulting spectra to a minimum of 5 counts per channel, and fitted those spectra with more than 50 net counts in total with an absorbed
blackbody model (TBABS*BBODYRAD in XSPEC, with the column density frozen at the value derived from the persistent emission; $3\times 10^{22} \,{\rm cm}^{-2}$).

We find a slow decay in temperature along the burst decay (``cooling tail''), from $\sim$2.9~keV to $\sim$1.4~keV, identifying this
unequivocally as a thermonuclear event. The burst bolometric peak
luminosity was $(10\pm4)\times 10^{37}\,{\rm erg\,s}^{-1}$, the apparent emitting radius
between 3 and 5 km (without color or redshift corrections), and the
total radiated energy about $1.0\times10^{39}\,{\rm erg}$ (see Figure~\ref{fig:burst}).
The persistent (0.5-10 keV) luminosity during the observation where
the burst occurred was $(5.6\pm0.1)\times 10^{36}\,{\rm erg\,s}^{-1}$ (about 5\% of the Eddington
limit for a bolometric correction factor of 2 and $L_{\rm Edd}=2.5\times10^{38}\,{\rm erg\,s}^{-1}$).

The peak of the 2015 outburst occurred on Apr 22, at
about ten times higher $L_X$, i.e., not far from 50\% $L_{\rm Edd}$ (for a $1.4 M_\odot$ NS). Despite good
Swift coverage ($\sim5$ ksec) of the following two weeks, when $L_X$ dropped by
about a factor 2, no other bursts were detected. 
This burst behaviour resembles that of
most thermonuclear bursters, where bursts virtually disappear at mass
accretion rates above 10\% Eddington. 

A second burster in Terzan 5, IGR 17480-2446, displays drastically different behaviour, namely a copious number of thermonuclear bursts at mass accretion rates between 10\% and 50\%
Eddington \citep{lin12}. The atypical behaviour in IGR J17480-2446 has been attributed to its slow (11 Hz) spin \citep{cav11,lin12}. Under this interpretation, the typical bursting behaviour of EXO 1745-248 would imply that it contains a rapidly rotating neutron star
($\sim200-600$ Hz), like most low-mass NSXBs. 

\cite{gallow08} define a burst timescale as $\tau = E_{\rm Burst}/F_{\rm Peak}$, where $E_{\rm Burst}$ is the total fluence during the burst and $F_{\rm Peak}$ is the peak flux of the burst. Following their definition, we find a burst timescale of $\approx 22$ s for EXO 1745$-$248. The 21 bursts seen by RXTE early in the 2000 outburst of EXO 1745$-$248 showed long burst durations ($\tau\sim25$ s) and other characteristics of H burning. However, two bursts seen later in the outburst were shorter ($\tau\sim10$ s), suggesting pure He (the explanation of this change in behaviour is not clear; \citealt{gallow08}). Therefore, we conclude that the measured timescale of this burst indicates the donor is likely hydrogen-rich.

\section{Conclusions}
In this paper, we present the results of our observations of the Terzan 5 NSXB EXO 1745$-$248 during its 2015 outburst at radio and X-ray frequencies, with the VLA, ATCA, and Swift XRT. Our (near-) simultaneous radio and X-ray measurements, all taken during the hard accretion state,  allow us to construct and fit the radio/X-ray correlation for this source ($L_R\propto L_X^\beta$; $\beta$ represents the disc-jet coupling index), which links the accretion flow to the relativistic jet in XBs. 
In contrast to the multiple BHXBs with a measured correlation, only two NSXBs have a measured radio/X-ray correlation,  Aql X-1 ($L_R\propto L_X^{0.76}$) and 4U 1728$-$34 ($L_R\propto L_X^{1.5}$). Additionally, an ensemble of tMSPs has been shown to follow a correlation, $L_R\propto L_X^{0.7}$, much more consistent with BHXBs.  As such, more measurements from NSXBs are needed to disentangle the different correlations. 
This work marks the third NSXB where the radio/X-ray correlation is measured in a single source, and the first where the distance is well known.

To fit the radio/X-ray correlation in EXO 1745$-$248 we developed a new MCMC based technique.
We find a best fit normalization and disc-jet coupling index for the radio/X-ray correlation in EXO 1745$-$248 of $\xi =0.94^{+0.14}_{-0.13}$ and $\beta=1.68^{+0.10}_{-0.09}$, respectively, where ($\log L_R-\log L_{R,c}) =\log\xi + \beta (\log L_X-\log L_{X,c})$, with centering values $L_{R,c}=3.89\times10^{28}\, {\rm erg\,s}^{-1}$ and $L_{X,c}=8.38\times10^{36}\, {\rm erg\,s}^{-1}$.

This disc-jet coupling index is consistent with what we would expect for a compact jet coupled to a radiatively efficient accretion flow (presumably due to the NSs surface), rather than a radiatively inefficient flow (as thought to exist in most BHXBs and possibly tMSPs). Empirically this index is consistent with the index for NSXB 4U 1728$-$34, but inconsistent with our measured index for NSXB Aql X-1. Therefore, a similar radio/X-ray correlation in the hard accretion state does not appear to hold across all three NSXBs measured so far, as it does in the BHXB population. However, all three NSXB correlations are measured over a smaller lever arm in X-ray luminosity ($\sim 1$ dex) when compared to BHXBs. 

Notably, we find that EXO 1745$-$248 is much more radio faint when compared to 4U 1728$-$34 and Aql X-1, where neither distance, mass, or spin considerations alone appear to be able to account for the discrepancy.

Finally, we detected an X-ray burst during this outburst. Through performing time-resolved spectral analysis, we find evidence of cooling during the decay of this burst and that the burst timescale is consistent with hydrogen burning, suggesting that this was a hydrogen Type-I X-ray burst. 

\section*{Acknowledgements}

AJT would like to thank Erik Rosolowsky for helpful discussions regarding MCMC implementation.
AJT, GRS, and COH are supported by NSERC Discovery Grants. JCAMJ is the recipient of an Australian Research Council Future Fellowship (FT140101082). ML was supported by the Spanish Ministry of Economy and Competitiveness under the grant AYA2013-42627. ET acknowledges support from NSF grant AST-1412549. JS acknowledges support via NSF grant AST-1308124. ND acknowledges support via an NWO/Vidi grant and an EU Marie Curie Intra-European fellowship under contract no. FP-PEOPLE-2013-IEF-627148. DA acknowledges support from the Royal Society. AP acknowledges support from an NWO Vidi fellowship. RW is supported by a NWO Top Grant, Module 1.
The National Radio Astronomy Observatory is a facility of the National Science Foundation operated under cooperative agreement by Associated Universities, Inc. This research has made use of the following data and software packages: Swift BAT transient monitor results provided by the Swift BAT team, and the Swift XRT Data Analysis Software (XRTDAS) developed under the responsibility of the ASI Science Data Center (ASDC), Italy.
We acknowledge extensive use of the ADS and arXiv.



\bibliography{ABrefList} 




\bsp	
\label{lastpage}
\end{document}